\documentclass[natbib]{svjour3}                     
\smartqed  
\usepackage{graphicx}
 \usepackage{aps-bibstyle}  
\begin{document}

\title{Testing the Stability of the Fine Structure Constant in the
Laboratory
}


\author{N.~Kolachevsky$^{1,2}$, A.~Matveev$^{1,2}$, J.~Alnis$^{1}$,
C.~Parthey$^1$, T.~Steinmetz$^{1,3}$, T.~Wilken$^{1}$,
R.~Holzwarth$^{3}$, Th.~Udem$^{1}$, T.W.~H\"ansch$^{1,4}$}


\institute{$^1$ Max-Planck Institut f\"ur Quantenoptik, Hans-Kopfermann-Str 1, 85748 Garching, Germany \\
           $^2$ P.N.~Lebedev Physical Institute, Leninsky prosp. 53, 119991 Moscow, Russia\\
           $^3$ MenloSystems GmbH, Am Klopferspitz 19, 82152 Martinsried, Germany\\
           $^4$ Ludwig-Maximilians-Universit\"at, Munich, Germany\\
           \email{kolik@lebedev.ru}}

\date{Received: date / Accepted: date}

\maketitle

\begin{abstract}
In this review we discuss the progress of the past decade in
testing for a possible temporal variation of the fine structure
constant $\alpha$. Advances in atomic sample preparation, laser
spectroscopy and optical frequency measurements led to rapid
reduction of measurement uncertainties. Eventually laboratory
tests became the most sensitive tool to detect a possible
variation of $\alpha$ at the present epoch. We explain the methods
and technologies that helped make this possible.
\end{abstract}

\section{Introduction}
\label{kolachevskyintro}

The fine structure constant $\alpha=e^2/\hbar c$ is a
dimensionless parameter that measures the strength of all
electromagnetic interactions. As such it appears in a large
variety of phenomena such as forces between charged objects that
in turn determine the structure of atoms and molecules. Further
examples are the propagation of electromagnetic waves, chemical
reactions or even macroscopic phenomena like friction. The value
of the fine structure constant can be thought of as the
electromagnetic force between two electrons at a distance of one
meter measured in units where the speed of light $c$ and Planck's
constant $\hbar$ are set to unity. The fine structure constant is
used as an expansion parameter in the quantum theory of
electromagnetic interactions, Quantum Electrodynamics (QED) which
is one the most successful theories in all physics.

Unfortunately neither this theory nor any other known theory makes
any prediction on the value of the fine structure constant which
is determined experimentally to $\alpha \approx 1/137$. Unlike
many other dimensionless numbers that we find in nature, such as
the number of particles in the Universe, the fine structure
constant represents a {\emph small} number. This fact has led
P.A.M.~Dirac to formulate his ``large number
hypothesis''~\cite{KolachevskyDirac1937} in 1937, where he
constructed small dimensionless numbers from known physical
constants assuming these are the fundamental parameters. One of
these small numbers is the age of the universe in atomic units
divided by the electromagnetic force between an electron and a
proton measured in units of their gravitational force and was
believed to be $\approx 3$ in 1937. Following this hypothesis, the
gravitational constant $G$ or any other constant that appears in
the construction of these small numbers should vary in time as the
Universe expands. This was the first alternative theory assuming
time-dependent coupling constants after Einstein's General
Relativity. Indeed, there is no theory yet that predicts the value
of $\alpha$ to be stable or drifting so that there is no reason to
expect one or the other behaviour. Even though Dirac's estimated
drift rate of $\alpha$ has been ruled out by repeated
measurements, the general possibility of ``variable constants''
remains open.

Modern theories that go beyond the large number hypnosis which
allow for the drift of fundamental constant rely on coupling
between gravitation and other fundamental interactions. Attempts
to unify gravity with electromagnetic, weak and strong
interactions encounter severe difficulties. To build such a
``theory of everything'' it seems that one has to extend the
number of dimensions of our usual space-time world. String
theories may allow for temporal and spatial variation of the
coupling constants that could be associated with cosmic dynamics.
Some possible mechanisms that lead to a drift or spatial
variations of the fundamental constants are discussed in
literature~\cite{KolachevskyTaylor1988,KolachevskyDamour2002,KolachevskyFlambaum2007a,KolachevskyFlambaum2007}.
As of now there is no sufficient theoretical evidence to make any
well-grounded prediction of the size of such variations. The
effect, if existing at all, should be extremely small since the
gravitational interaction seems to be almost decoupled at the
low-energy limit. For this reason experimental research is the
appropriate way to probe this type of new physics that goes beyond
the standard model.

The basic principle of every experimental search for a time
variation of fundamental constants is the measurement of a
physical quantity $\Phi(\gamma_1, \cdots, \gamma_K, t)$, which is
a function of several fundamental constants $\gamma_i$, at times
$t_1$ and $t_2$, separated by the interval $\Delta t=t_1-t_2$. If
$\Phi(\gamma_1, \cdots, \gamma_K, t)$ is a function of more than
one constant ($K>1$) it is not possible to derive separate values
for $\Delta \gamma_i$ even if the dependence of $\Phi$ on
$\gamma_i$ is known. In addition if $\Phi$ is not a dimensionless
number it will have to be compared to some reference. The
dimensionless ratio resulting from such a measurement then
contains possible variations of $\Phi$ and of the reference.
However, repeated measurements on several physical quantities
$\Phi_j$ with $j=1,\cdots,N$ and ($N \geq K$) or assumptions on
restrictions or mutual correlations of the constants or their
drift rates may be used to derive all $\Delta \gamma_i$ involved.
Lacking any accepted theory of the variation of fundamental
constants one would prefer to draw conclusions with the smallest
possible set of assumptions.

Concerning the time interval $\Delta t$, there are two extreme
classes of experiments: (i) astronomical or geological
observations and (ii) high precision laboratory measurements. The
investigation of absorption or emission lines of distant galaxies
back illuminated by the white light of quasars at even larger
distances takes advantage of the extremely long look back time of
up to $10^{10}$ years. In contrast to that, laboratory frequency
comparisons are restricted to short time intervals of a few years
but can be as sensitive if precision measurements with
uncertainties of better than $10^{-15}$ are performed. Currently
such a low uncertainty can only be realized by frequency
measurements in the radio or optical domain. For this reason
frequency comparisons of atomic, molecular or ionic transitions
are used. The important advantages of laboratory experiments are:
The variety of different systems that may be tested, the
possibility to change parameters of the experiments in order to
control systematic effects and the straightforward determination
of the drift rates from the measured values. Modern precision
frequency measurements deliver information about the stability of
the to-date values of the constants, which can only be tested with
laboratory measurements. At the same time only non-laboratory
methods are sensitive to processes that occurred in the early
Universe, which may be much larger than at present times. As both
classes of experiments (i) and (ii) probe $\Delta \gamma_i$ at
different epochs, they supplement each other to get a more
detailed view of the possible time variation of fundamental
constants.

In 2000 J.K.~Webb and co-workers introduced the many-multiplet
method~\cite{KolachevskyWebb2001}, which is an extension of the
alkali-doublet
method~\cite{KolachevskySav56,KolachevskyMinkowski56}, to extract
the value of the fine structure constant from quasi-stellar object
(QSO) absorption spectra. As the detected absorption lines emerged
billions of light years away they conserve the value of $\alpha$
over that period of time. Application of the many-multiplet method
to KECK/HIRES QSO data indicated that $\alpha$ was smaller by
$\Delta\alpha/\alpha=(5.4 \pm 1.2) \times 10^{-6}$ about $10^{10}$
years ago~\cite{KolachevskyWebb2001}. This $5\sigma$ deviation
from a non drifting value stimulated further investigation in the
field. In 2003-2004 another set of astrophysical data obtained by
the Very Large Telescope was analyzed independently using the same
approach~\cite{KolachevskyChand2004,KolachevskyQuast2004}. The
conclusion was that $\alpha$ was stable within
$|\Delta\alpha/\alpha|<10^{-6}$ in the past. Meanwhile M.T.~Murphy
and co-workers pointed out some possible flaws in the data
evaluation~\cite{KolachevskyMurphy2008} so that astrophysical data
remain contradictory (see also~\cite{KolachevskyVarshalovich}).
Unlike laboratory measurements astrophysical data analysis
strongly relies on cosmological evolution, i.e. expansion,
isotopic abundances, magnetic field distribution, etc. which are
also debated.

\begin{figure}[t!]
\begin{center}
\includegraphics[width=0.5\textwidth]{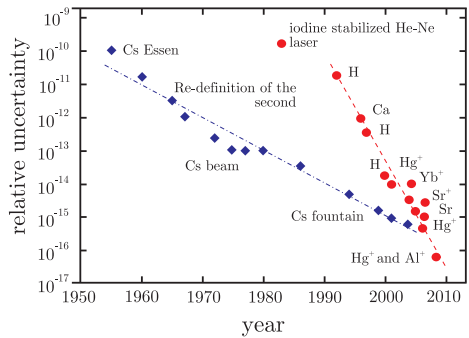}
\caption{Uncertainties of the microwave Cs ground state hyperfine
splitting (diamonds) and optical frequency standards based on a
variety of atoms and ions (circles).}
\label{kolachevskyfig1}       
\end{center}
\end{figure}
Laboratory frequency measurements have become competitive very
recently in terms of sensitivity to a possible variation of
$\alpha$ in the present epoch. Figure~\ref{kolachevskyfig1}
summarizes the progress achieved during the last decades in the
field of optical frequency measurements (only a few are selected
for the plot). For comparison, the progress in microwave frequency
standards used for realization of the SI second is shown in the
same plot. Improvements of the last years have been due to new
ultra-cold atomic samples, laser stabilization techniques as well
as breakthroughs in optical frequency measurements so that
relative uncertainties in the optical domain are approaching
$10^{-17}$.

\begin{figure*}[t!]
\begin{center}
\includegraphics[width=0.97\textwidth]{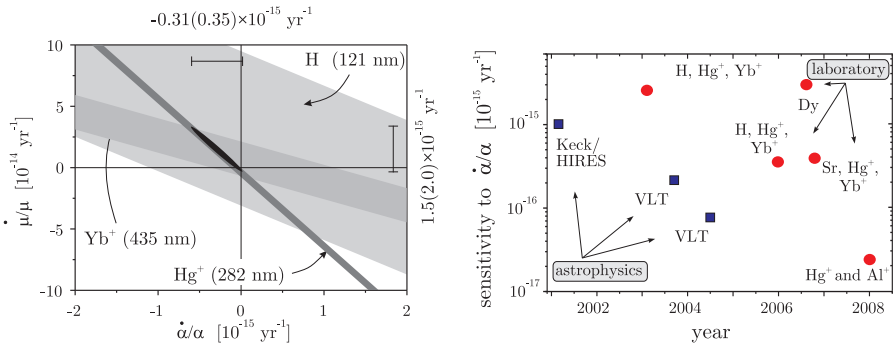}
\caption{(left) ---  Illustration of the laboratory method for
detecting possible variations of the fine structure constant
$\alpha$. The method is based on combinations of absolute
frequency measurements in different atomic
systems~\cite{KolachevskyPeik2004,KolachevskyFischer2004,KolachevskyFortier2007}.
(right)
--- Comparison of sensitivities of astrophysical and laboratory
methods for a presumed linear drift of $\alpha$. }
\label{kolachevskyfig2}       
\end{center}
\end{figure*}

With the introduction of frequency combs (see section
\ref{kolachsub12}) high-precision optical frequency measurements
became a routine procedure, readily available for a broad
scientific
community~\cite{KolachevskyHolzwarth2000,KolachevskyUdem2002}.
Repeated frequency measurements of some atomic transitions allowed
to tighten the upper limit for the variation of frequency ratios.
The latter can be used in a variety of fundamental tests,
including the search for the variation of $\alpha$. Optical
frequency measurements from between 2000 and 2003 in
ytterbium~\cite{KolachevskyPeik2004} and
mercury~\cite{KolachevskyBize2003} ions as well as in atomic
hydrogen~\cite{KolachevskyFischer2004} allowed impose a
model-independent restriction of
$\dot\alpha/\alpha=(-0.9\pm2.9)\times 10^{-15}\,\textrm{yr}^{-1}$
(see fig.\ref{kolachevskyfig2},\,left). The sensitivity of this
test was already competitive to the sensitivity of the KECK/HIRES
data analysis~\cite{KolachevskyWebb2001}
(fig.\ref{kolachevskyfig2},\,right) which was the most sensitive
analysis from astrophysical observations at that time.

The laboratory method is based on optical frequency measurements
and allowed increasing the sensitivity of probing the possible
$alpha$ variation by an order of magnitude already in
2005~\cite{KolachevskyPeik2006,KolachevskyBlatt2008}. Thus far the
lowest limited on the present drift rate of $\alpha$ has been
obtained by T.~Rosenbad and co-workers at NIST (USA) by direct
comparison of optical clock transitions in mercury and aluminum
ions via a frequency comb~\cite{KolachevskyRosenband2008}. Their
result reads $\dot\alpha/\alpha=(-1.6\pm2.3)\times 10^{-17}
\,\textrm{yr}^{-1}$ and is an order of magnitude more accurate
than astrophysical observations albeit at a different epoch.

In what follows we present modern techniques used for spectroscopy
and frequency measurement of narrow optical transitions
(section~\ref{kolachevskysec1}), discuss the model-independent
laboratory method for restricting the variation of $\alpha$
(section~\ref{kolachevskysec2}) and point out some perspectives
opened by optical frequency metrology for astrophysics
(section~\ref{kolachevskysec3}).

\section{Precision optical spectroscopy and optical frequency measurements}
\label{kolachevskysec1}

\begin{figure}[t!]
\begin{center}
\includegraphics[width=0.66\textwidth]{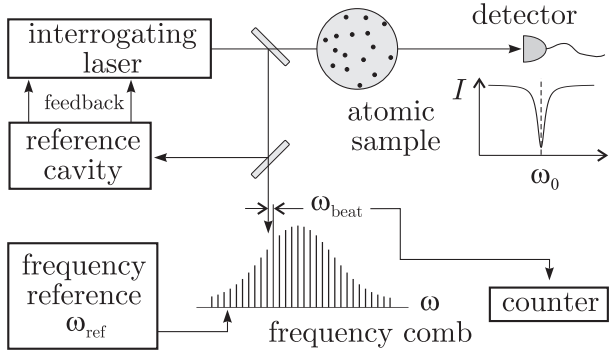}
\caption{Setup for the measurement an optical transition frequency
in an atomic sample with the help of an optical frequency comb.}
\label{kolachevskyfig3}
 \end{center}
\end{figure}

The principle of modern optical frequency measurement is presented
in fig.~\ref{kolachevskyfig3}. A laser is tuned to the wavelength
of a narrow metrological transition (usually referred to as a
``clock transition'') in an atomic, ionic or molecular sample.
Most commonly, the laser frequency is stabilized by active
feedback to a transmission peak of a well isolated optical cavity
(``reference cavity'') which allows to achieve sub-hertz spectral
line width of the interrogating laser. Some recent advances in
laser stabilization technique will be described in
section~\ref{kolachsub11}. The laser frequency is then scanned
across the transition which allows to find the line center
$\omega_0$ using an appropriate line shape model. The measured
transition quality factor can reach $10^{15}$ which provides
extremely high resolution. To obtain the transition frequency the
beat note $\omega_\mathrm{beat}$  between the laser and one of the
modes of the stabilized frequency comb is measured with the help
of a frequency counter. Details of this type of measurement are
presented in section~\ref{kolachsub12}. If the comb is stabilized
to a primary frequency reference (i.e. a Cs atomic clock), the
measurement presented in fig.~\ref{kolachevskyfig3} will yield the
{\it absolute} frequency of the optical transition. Absolute
frequency measurements allow a comparison of different results
obtained at laboratories all over the world. An example of such a
comparison is given in section~\ref{kolachsubsec3}. On the other
hand, if the comb is stabilized with the help of some other
reference, which can be e.g. another optical frequency, the
measurement will yield the ratio $\omega_0/\omega_\mathrm{ref}$.
One can thus compare transition frequencies in different atomic
samples avoiding time-consuming absolute frequency measurements
and avoiding the additional uncertainty introduced by primary
frequency standards (fig.~\ref{kolachevskyfig1}).

\subsection{Laser stabilization} \label{kolachsub11}

Interrogation of a clock transition in an atomic sample requires a
narrow-band laser source. Due to vibrations, fluctuations of the
pump intensity as well as some intrinsic noise sources (e.g. the
excessive Henry phase noise in semiconductor
lasers~\cite{KolachevskyHenry1982}), the typical laser line width
turns out to be many orders of magnitude broader than the
Schawlow-Townes limit~\cite{KolachevskySchawlow1958}. A passive
isolation of the laser resonator itself is not sufficient to
suppress these noise sources so that active stabilization to an
external reference cavity is implemented.

The reference cavity should be well isolated from the environment
by placing it in a separate vibrationally and thermally stabilized
vacuum chamber. If the laser frequency is stabilized to the
transmission peak of such a cavity, e.g. by means of the
Pound-Drever-Hall technique~\cite{KolachevskyPound1983}, the laser
frequency fluctuations $\delta \nu$ will be directly coupled to
the fluctuations of the cavity length $\delta l$ according to
$\delta \nu/\nu=\delta l/l$, where $\nu$ is the laser frequency,
and $l$ is the cavity length. If one desires to achieve $\delta
\nu=$1\,Hz using a cavity of $l=10$\,cm, the distance between the
mirrors should remain constant at the level of $10^{-16}$\,m,
which is a fraction of the proton radius!

The first demonstration of a sub-Hz laser line width by B.C.~Young
and co-workers in 1999 made use of a heavy optical bench suspended
with rubber tubes from the lab ceiling~\cite{KolachevskyYoung1999}
for vibration isolation. In the meantime cavity designs and
mountings emerged where deformations due to vibrations do not
change the critical length that separates the
mirrors~\cite{KolachevskyNotcutt2006,KolachevskyNazarova2006}.
This has not only led to much more compact set-ups but also to a
number of laser sources successfully stabilized to sub-hertz
level~\cite{KolachevskyStoehr2006,KolachevskyLudlow2007}. In our
laboratory we use vertically mounted Fabry-P\'erot (FP) cavities,
with a spacer design from A.D.~Ludlow and
co-workers~\cite{KolachevskyLudlow2007} and reach 40\,dB
suppression of vertical vibration
sensitivity~\cite{KolachevskyAlnis2008}.

\begin{figure*}[t!]
\begin{center}
\includegraphics[width=0.95\textwidth]{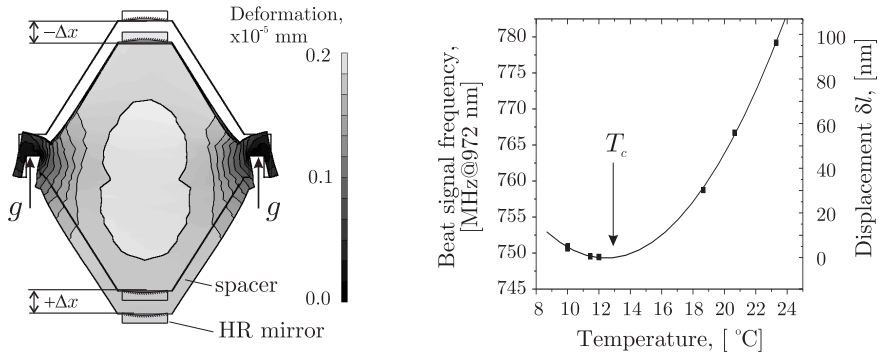}
\caption{ (left) --- Deformations of a vertically mounted
Fabry-P\'erot (FP) cavity under the influence of vertical
acceleration of $1g$ determined with finite elements analysis. The
mounting is such that the compression of upper part $-\Delta x$ is
compensated by the stretch of the lower part $+\Delta x$
maintaining the critical distance of the mirrors. (right)
--- Determination of the zero-expansion temperature $T_c$ by
measuring the beat note frequency between the FP cavity and the
laser stabilized to the second FP cavity maintained at a constant
temperature. At $T_c$ the cavity length $l$ reaches the minimum. }
\label{kolachevskyfig4}
\end{center}
\end{figure*}

The principle of such a cavity mounting is shown in the left hand
side of fig.~\ref{kolachevskyfig4}. The cavity spacer is suspended
at its midplane such that the influence of vibration induced
vertical acceleration to the mirror separation is significantly
suppressed. This makes the setup virtually immune for vertical
vibrations that are more difficult to suppress than horizontal
ones. If such a cavity is placed on a vibration-isolated platform,
the acoustic and seismic vibrations from the environment have
virtually no detectable influence on the cavity frequency.

Another issue that affects the stability for averaging times
larger than several seconds is the dimensional stability due to
temperature variations. Certain glass ceramics can be made with
very low thermal expansion and the one made by Corning is called
ultra low expansion glass (ULE). ULE is a titania-doped silicate
glass that has a specified thermal expansion minimum at some
temperature $T_c$ around room temperature according to
\begin{equation}\label{kolach_eq1}
\delta l/\l\sim 10^{-9}(T-T_c)^2\,,
\end{equation}
where $T$ is the cavity temperature. To reduce the quadratic
dependence the temperature of the material should be stabilized as
close as possible to $T_c$. Unfortunately the measured composite
thermal stationary point of the cavity resonance frequency usually
ends up below room temperature~\cite{KolachevskyFox2008}
(fig.~\ref{kolachevskyfig4}, right). This poses a problem because
cooling of the vacuum chamber is more difficult than heating as
water condensation on the windows prevents the coupling of laser
light into the chamber. The problem was solved by cooling the FP
cavity directly in the vacuum chamber by Peltier
elements~\cite{KolachevskyAlnis2008}.

\begin{figure}[t!]
\begin{center}
\includegraphics[width=0.95\textwidth]{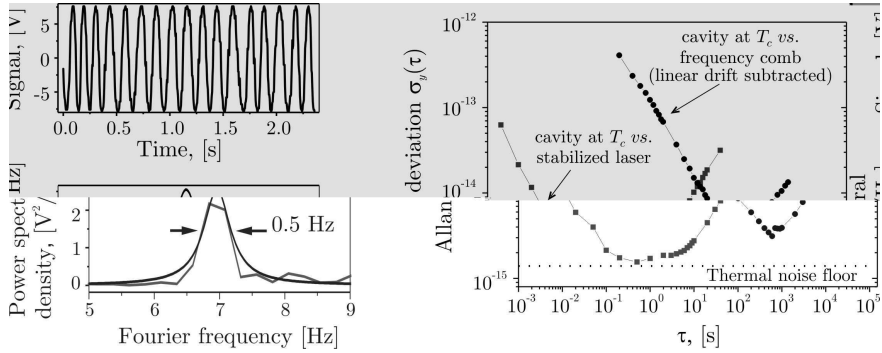}
\caption{ (left) --- Oscillogram of the beat note signal between
two diode lasers locked to two independent vertically mounted FP
cavities, its Fourier transformation and Lorentzian fit. (right)
--- Allan deviation of the beat note signal from the left part of
the figure (squares). The cavity kept at $T_c$ is extremely stable
as can be seen from its absolute frequency measurement performed
with an optical frequency comb (see next section) which was
referenced to a hydrogen maser (circles).} \label{kolachevskyfig5}
\end{center}
\end{figure}

This type of vibration- and thermal compensation allows to set up
extremely stable and compact laser sources. For example, the diode
lasers operating at 972\,nm designed for two-photon spectroscopy
of the $1S$\,-\,$2S$ clock transition in atomic hydrogen are
characterized as shown in fig.~\ref{kolachevskyfig5}. Two nearly
identical systems have been built in our lab with one FP cavity
maintained near $T_c$ of its resonance frequency, while the other
stabilized to a temperature 25\,$^\circ$C above that point. Both
lasers demonstrate excellent short-time stability (up to 10\,s)
approaching the thermal noise limit of $10^{-15}$ which is set by
the Brownian motion of the mirror surfaces. Concerning the
long-term stability, the FP cavity at $T_c$ demonstrates a much
better performance since it is much less influenced by ambient
temperature fluctuations. It possesses a linear drift of about
$+50$\,mHz/s mainly caused by ULE aging, while its resonance
frequency deviates from that linear drift by only $\pm10$\,Hz on a
time scale of 10\,hrs.

Excellent spectral characteristics and small size of the setup
allow to use such lasers in the most demanding applications of
frequency metrology. Possible routes to overcome the thermal noise
limit are currently being discussed and may result in further
improvements of the laser spectral purity.

\subsection{Ultra-short pulse lasers and frequency combs}
\label{kolachsub12}

Frequency can be measured with by far the highest precision of all
physical quantities. In the radio frequency domain (say up to
100~GHz), frequency counters have existed for a long time. Almost
any of the most precise measurements in physics have been
performed with such a counter that uses an atomic clock as a time
base. To extend this accurate technique to higher frequencies, so
called harmonic frequency chains have been constructed since the
late 1960ies~\cite{KolachevskyHocker1967,KolachevskyEvenson1973}.
Because of the large number of steps necessary to build a long
harmonic frequency chain, it was not before 1995 when visible
laser light was first referenced phase coherently to a cesium
atomic clock using this method~\cite{KolachevskySchnatz1996}.

The disadvantage of these harmonic frequency chains was not only
that they could easily fill several large laser laboratories at
once, but that they could be used to measure a single optical
frequency only. Even though mode locked lasers for optical
frequency measurements have been used in rudimentary form in the
late 1970ies~\cite{KolachevskyEckstein1978}, this method became
only practical with the advent of femtosecond (fs) mode locked
lasers. Such a laser necessarily emits a very broad spectrum,
comparable in width to the optical carrier frequency.

In the frequency domain a train of short pulses from a femtosecond
mode locked laser is the result of a phase coherent superposition
of many continuous wave (cw) longitudinal cavity modes. These
modes at $\omega_n$ form a series of frequency spikes that is
called a frequency comb. As has been shown, the modes are
remarkably uniform, i.e. the separation between adjacent modes is
constant across the frequency
comb~\cite{KolachevskyHolzwarth2000,KolachevskyUdem1999a,KolachevskyDiddams2002,KolachevskyMa2004}.
This strictly regular arrangement is the most important feature
used for optical frequency measurement and may be expressed as:
\begin{equation}
  \omega_n = n \omega_r + \omega_{CE}\,.
  \label{kolach_eq2}
\end{equation}
Here the mode number $n$ of some $10^5$ may be enumerated such
that the frequency offset $\omega_{CE}$ lies in between $0$ and
$\omega_r=2\pi/T$. The mode spacing is thereby identified with the
pulse repetition rate, i.e. the inverse pulse repetition time $T$.
With the help of that equation two radio frequencies $\omega_r$
and $\omega_{CE}$ are linked to the optical frequencies $\omega_n$
of the laser. For this reason mode locked lasers are capable to
replace the harmonic frequency chains of the past.

To derive the frequency comb
properties~\cite{KolachevskyReichert1999} as detailed by
(\ref{kolach_eq2}), it is useful to consider the electric field
$E(t)$ of the emitted pulse train. We assume that the electric
field $E(t)$, measured for example at the lasers output coupling
mirror, can be written as the product of a periodic envelope
function $A(t)$ and a carrier wave $C(t)$:
\begin{equation}
  E(t) = A(t)C(t) + c.c.\,.
  \label{kolach_eq3}
\end{equation}
The envelope function defines the pulse repetition time
$T=2\pi/\omega_r$ by demanding $A(t)=A(t-T)$. The only thing about
dispersion that should be added for this description, is that
there might be a difference between the group velocity and the
phase velocity inside the laser cavity. This will shift the
carrier with respect to the envelope by a certain amount after
each round trip. The electric field is therefore in general not
periodic with $T$. To obtain the spectrum of $E(t)$ the Fourier
integral has to be calculated:
\begin{equation}\label{kolach_eq4}
  \tilde{E}(\omega) = \int_{-\infty}^{+\infty} E(t) e^{i\omega t}
  dt\,.
\end{equation}
Separate Fourier transforms of $A(t)$ and $C(t)$ are given by:
\begin{equation}\label{kolach_eq5}
  \tilde{A}(\omega) = \sum_{n=-\infty}^{+\infty} \delta \left(\omega - n \omega_r \right) \tilde{A}_n
  \hspace{1cm} \mbox{and} \hspace{1cm}
  \tilde{C}(\omega) = \int_{-\infty}^{+\infty} C(t) e^{i\omega t}
  dt\,.
\end{equation}
A periodic frequency chirp imposed on the pulses is accounted for
by allowing a complex envelope function $A(t)$. Thus the
``carrier'' $C(t)$ is defined to be whatever part of the electric
field that is non-periodic with $T$. The convolution theorem
allows us to calculate the Fourier transform of $E(t)$ from
$\tilde{A}(\omega)$ and $\tilde{C}(\omega)$:
\begin{equation}
  \tilde{E}(\omega) = \frac{1}{2\pi} \int_{-\infty}^{+\infty}
  \tilde{A}(\omega') \tilde{C}(\omega - \omega') d \omega'  + c.c.
  =\frac{1}{2\pi} \sum_{n=-\infty}^{+\infty} \tilde{A}_n \tilde{C}
  \left( \omega - n \omega_r \right) + c.c.\,.
  \label{kolach_eq6}
\end{equation}

The sum represents a periodic spectrum in frequency space. If the
spectral width of the carrier wave $\Delta \omega_c$ is much
smaller than the mode separation $\omega_r$, it represents a
regularly spaced comb of laser modes just like (\ref{kolach_eq2}),
with identical spectral line shapes. If $\tilde{C}(\omega)$ is
centered at say $\omega_c$, then the comb is shifted by $\omega_c$
from containing only exact harmonics of $\omega_r$. The
frequencies of the mode members are calculated from the mode
number
$n$~\cite{KolachevskyUdem2002,KolachevskyEckstein1978,KolachevskyReichert1999}:
\begin{equation}
  \omega_n = n \omega_r + \omega_{c}\,.
  \label{kolach_eq7}
\end{equation}
The measurement of the $\omega_c$ as described below (see
also~\cite{KolachevskyHolzwarth2000,KolachevskyUdem2002,KolachevskyUdem1999a,KolachevskyReichert1999,KolachevskyDiddams2000})
usually yields a value modulo $\omega_r$, so that renumbering the
modes will restrict the offset frequency to smaller values than
the repetition frequency and (\ref{kolach_eq2}) and
(\ref{kolach_eq7}) are identical.

\begin{figure}[t!]
\begin{center}
\includegraphics[width=0.6\textwidth]{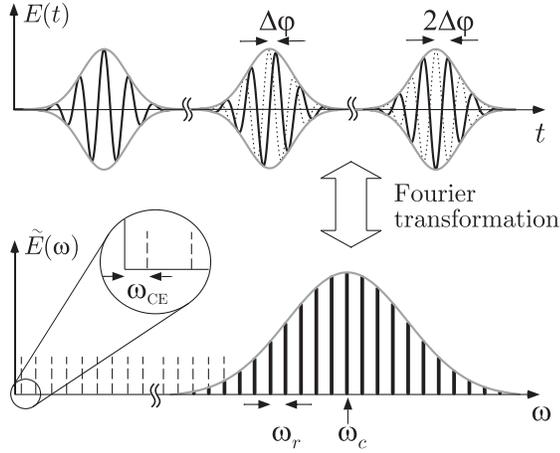}
\caption{{Consecutive un-chirped pulses ($A(t)$ is real) with
carrier frequency $\omega_c$ and the corresponding spectrum (not
to scale). Because the carrier propagates with a different
velocity within the laser cavity than the envelope (with phase-
and group velocity respectively), the electric field does not
repeat itself after one round trip. A pulse-to-pulse phase shift
$\Delta \varphi$ results in an offset frequency of
$\omega_{CE}=\Delta\varphi/T$. The mode spacing is given by the
repetition rate $\omega_r$. The width of the spectral envelope is
given by the inverse pulse duration up to a factor of order unity
that depends on the pulse shape.}} \label{kolachevskyfig6}
\end{center}
\end{figure}

If the carrier wave is monochromatic $C(t)=e^{-i\omega_c t - i
\varphi}$, its spectrum will be $\delta$-shaped and centered at
the carrier frequency $\omega_c$. The individual modes are also
$\delta$-functions $\tilde{C}(\omega)= \delta (\omega-\omega_c)
e^{-i\varphi}$. The frequency offset (\ref{kolach_eq7}) is
identified with the carrier frequency. According to
(\ref{kolach_eq3}) each round trip will shift the carrier wave
with respect to the envelope by $\Delta
\varphi=\arg(C(t-T))-\arg(C(t))=\omega_c T$ so that the frequency
offset may also be identified by $\omega_{CE}=\Delta
\varphi/T$~\cite{KolachevskyUdem2002,KolachevskyEckstein1978,KolachevskyReichert1999}.
In a typical laser cavity this pulse-to-pulse carrier-envelope
phase shift is much larger than $2\pi$, but measurements usually
yield a value modulo $2\pi$. The restriction $0 \leq \Delta
\varphi \leq 2\pi$ is synonymous with the restriction $0 \leq
\omega_{CE} \leq \omega_r$ introduced above.
Figure~\ref{kolachevskyfig6} sketches this situation in the time
domain for a chirp free pulse train.

\subsubsection{Extending the frequency comb}

The spectral width of a pulse train emitted by a fs laser can be
significantly broadened in a single mode
fiber~\cite{KolachevskyAgrawal2001} by self phase modulation.
Assuming a single mode carrier wave, a pulse that has propagated
the length $L$ acquires a self induced phase shift of
\begin{equation}\label{kolach_eq8}
  \Phi_{NL}(t)=-n_2I(t)\omega_c L/c\,,
\end{equation}
where the pulse intensity is given by $I(t)=\frac{1}{2} c
\varepsilon_0 \vert A(t)\vert^2$. For fused silica the non-linear
Kerr coefficient $n_2$ is comparatively small but almost
instantaneous even on the time scale of fs pulses. This means that
different parts of the pulse travel at different speed. The result
is a frequency chirp across the pulse without affecting its
duration. The pulse is no longer at the Fourier limit so that the
spectrum is much broader than the inverse pulse duration where the
extra frequencies are determined by the time derivative of the
self induced phase shift $\dot{\Phi}_{NL}(t)$. Therefore pure
self-phase modulation would modify the envelope function in
(\ref{kolach_eq3}) according to
\begin{equation}\label{kolach_eq9}
  A(t) \longrightarrow A(t)e^{i\Phi_{NL}(t)}.
\end{equation}
Because $\Phi_{NL}(t)$ has the same periodicity as $A(t)$ the comb
structure of the spectrum is maintained and the derivations
(\ref{kolach_eq6}) remain valid because periodicity of $A(t)$ was
the only assumption made. An optical fiber is most appropriate for
this process because it can maintain the necessary small focus
area over a virtually unlimited length. In practice, however,
other pulse reshaping mechanism, both linear and non-linear, are
present so that the above explanation might be too simple.

\begin{figure}[h]
\begin{center}
\includegraphics[width=7cm]{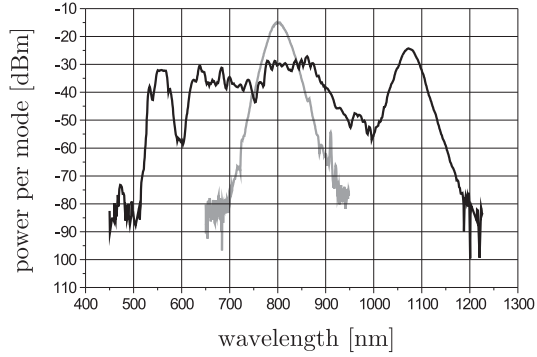}
\caption{{Power per mode of the frequency comb on a logarithmic
scale (0~dBm\,=\,1mW). The lighter 30~nm (14~THz at $-$3~dB) wide
spectrum displays the laser intensity and the darker octave
spanning spectrum (532~nm through 1064~nm) is observed after
spectral broadening in a 30~cm microstructured
fiber~\cite{KolachevskyKnight1996}. The laser was operated at
$\omega_r=2\pi \times 750$~MHz (modes not resolved) with 25~fs
pulse duration. An average power of 180~mW was coupled through the
microstructure fiber~\cite{KolachevskyHolzwarth2001}.}}
\label{kolachevskyfig7}
\end{center}
\end{figure}

A microstructured fiber uses an array of submicron-sized air holes
that surround the fiber core and run the length of a silica fiber
to obtain a desired effective
dispersion~\cite{KolachevskyKnight1996}. This can be used to
maintain the high peak power over an extended propagation length
and to significantly increase the spectral broadening. With these
fibers it became possible to broaden low peak power, high
repetition rate lasers to beyond one optical octave as shown in
fig.~\ref{kolachevskyfig7}.

Another class of frequency combs that can stay in lock for longer
times are fs fiber lasers~\cite{KolachevskyNelson1997}. The most
common type is the erbium doped fiber laser that emits within the
telecom band around 1550~nm. For this reason advanced and cheap
optical components are available to build such a laser. The mode
locking mechanism is similar to the Kerr lens method, except that
non-linear polarization rotation is used to favor the pulsed high
peak intensity operation. Up to a short free space section that
can be build very stable, these lasers have no adjustable parts.
Continuous stabilized operation for many
hours~\cite{KolachevskyKubina2005,KolachevskyAdler2004} has been
reported. The Max-Planck Institute f\"ur Quantenoptik in Garching
(Germany) operates a fiber based self referenced frequency comb
that stays locked without interruption for months.

\subsubsection{Self-referencing}

The measurement of $\omega_{CE}$ fixes the position of the whole
frequency comb and is called self-referencing. The method relies
on measuring the frequency gap between {\em different} harmonics
derived from the {\em same} laser or frequency comb. The simplest
approach is to fix the absolute position of the frequency comb by
measuring the gap between $\omega_n$ and $\omega_{2n}$ of modes
taken directly from the frequency
comb~\cite{KolachevskyHolzwarth2000,KolachevskyUdem2002,KolachevskyDiddams2000,KolachevskyReichert2000}.
In this case the carrier-envelope offset frequency $\omega_{CE}$
is directly produced by beating the frequency doubled\footnote{It
should be noted that this does not simply mean the doubling of
each individual mode, but the general sum frequencies generation
of all modes. Otherwise the mode spacing, and therefore the
repetition rate, would be doubled as well.} red wing of the comb
$2\omega_n$ with the blue side of the comb at $\omega_{2n}$:
$2\omega_n-\omega_{n'}=(2n-n') \omega_r +\omega_{CE}=\omega_{CE}$
where again the mode numbers $n$ and $n'$ are chosen such that
$(2n-n')=0$. This approach requires an octave spanning comb, i.e.
a bandwidth of 375~THz if centered at the titanium-sapphire gain
maximum at 800~nm.

\begin{figure}[ht]
\begin{center}
\includegraphics[width=\textwidth]{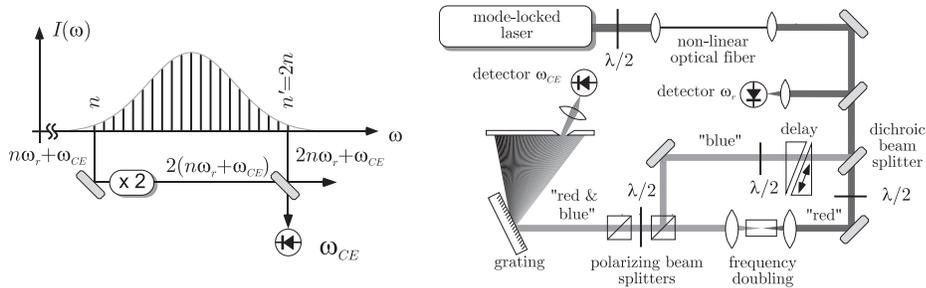}
\caption[]{{ (left) --- The principle of the $f-2f$ self
referencing relies on detecting a beat note at $\omega_{CE}$
between the frequency doubled ``red'' wing
$2(n\omega_r+\omega_{CE})$ of the frequency comb and the ``blue''
modes at $2n\omega_r+\omega_{CE}$. (right)
--- More detailed layout of the self referencing scheme. See text
for details.}} \label{kolachevskyfig8}
\end{center}
\end{figure}

Figure~\ref{kolachevskyfig8} sketches the $f-2f$ self referencing
method. The spectrum of a mode locked laser is first broadened to
more than one optical octave with an optical fiber. A broad band
$\lambda/2$ wave plate allows to choose the polarization with the
most efficient spectral broadening. After the fiber a dichroic
mirror separates the infrared (``red'') part from the green
(``blue''). The former is frequency doubled in a non-linear
crystal and reunited with the green part to create a wealth of
beat notes, all at $\omega_{CE}$. These beat notes emerge as
frequency difference between $2\omega_n-\omega_{2n}$ according to
(\ref{kolach_eq2}) for various values of $n$. The number of
contributing modes is given by the phase matching bandwidth
$\Delta \nu_{pm}$ of the doubling crystal and can easily exceed
1~THz.

As described, both degrees of freedom $\omega_r$ and $\omega_{CE}$
of the frequency comb can be measured up to a sign in
$\omega_{CE}$ that will be discussed below. For stabilization of
these frequencies, say relative to a radio frequency reference, it
is also necessary to control them. Again the repetition rate turns
out to be simpler. Mounting one of the laser's cavity mirrors on a
piezo electric transducer allows to control the pulse round trip
time. Controlling the carrier envelope frequency requires some
effort. Any laser parameter that has a different influence on the
cavity round trip phase delay and the cavity round trip group
delay may be used to change
$\omega_{CE}$~\cite{KolachevskyHaus2001}. Experimentally it turned
out that the energy of the pulse stored inside the mode locked
laser cavity has a strong influence on $\omega_{CE}$. To phase
lock the carrier envelope offset frequency $\omega_{CE}$, one can
therefore control the laser power through its energy source (pump
laser).

\subsubsection{Frequency conversions} \label{kolachevskysubsec223}

Given the above we conclude that the frequency comb may serve as a
frequency converter between the optical and radio frequency
domains allowing to perform the following phase coherent
operations:
\begin{itemize}
\item{convert a radio frequency into an optical frequency. In this
case both $\omega_r$ and $\omega_{CE}$ from (\ref{kolach_eq2}) are
directly locked to the radio frequency source.}
\item{convert an optical frequency into a radio frequency. In this
case the frequency of one of the comb modes $\omega_n$ is locked
to a clock laser while the carrier envelope frequency $\omega_{CE}$
is phase locked to $\omega_r$. The repetition rate will then be used
as the countable clock output.}
\item{convert an optical frequency to another optical frequency, i.e.
measuring optical frequency ratios. In this case the comb is stabilized
to one of the lasers as described in the second case, but instead of
measuring $\omega_r$ one measures the beat note frequency between
another laser and its closest comb mode $\omega_n'$.}

\end{itemize}

\subsection{Frequency measurement of the $1S$\,-\,$2S$ transition in atomic hydrogen}

One of the first optical frequency measurements performed with an
optical frequency comb was the $1S$\,-\,$2S$ transition frequency
in atomic hydrogen in our lab. During the last decades precision
spectroscopic experiments on hydrogen and deuterium atoms have
yielded new accurate values for the Rydberg
constant~\cite{KolachevskyBiraben2001}, the ground-state Lamb
shift~\cite{KolachevskyWeitz1994}, the deuteron structure
radius~\cite{KolachevskyHuber1998}, and the $2S$ hyperfine
structure~\cite{KolachevskyKolach2004a,KolachevskyKolach2004b}.
Accurate optical frequency measurements allow for sensitive tests
of quantum electrodynamics (QED), which are based on comparisons
between experimental values and results from corresponding QED
calculations (for review
see~\cite{KolachevskyEides2001,KolachevskyKarshen2002}).

To measure the frequency $\omega_L$ of the continuous wave (cw)
interrogation laser (486\,nm) that drives the $1S$\,--\,$2S$
transition, a beat note $\omega_b$ with a stabilized frequency
comb is generated (see fig.~\ref{kolachevskyfig8}). For this
purpose the beam of the cw laser is spatially overlapped with the
beam that contains the frequency comb and guided to a photo
detector. The frequency of the interrogation laser is then given
by
\begin{equation}\label{kolach_eq10}
  \omega_L=n \omega_r \pm \omega_{CE} \pm \omega_b\,.
\end{equation}
The signs can be determined by introducing small changes to
$\omega_r$ and $\omega_{CE}$ and observing the corresponding shift
in $\omega_b$. This uniquely fixes both signs if $\omega_L$ is
held fixed during this test. The mode number $n$ may be determined
by a coarse measurement say with a high-resolution wave meter, by
re-measuring with different $\omega_r$ or by comparison with
previous results of lower accuracy.

\begin{figure}[ht]
\begin{center}
\includegraphics[width=0.5\textwidth]{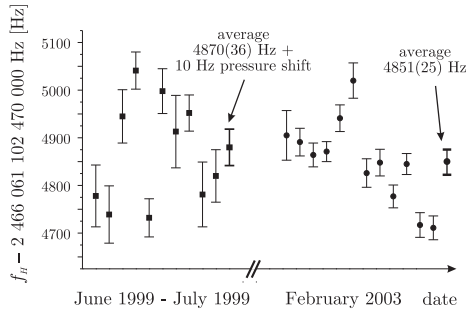}
\caption[]{{Absolute frequency measurements of the $1S$\,-\,$2S$
transition in atomic hydrogen.}} \label{kolachevskyfig9}
\end{center}
\end{figure}

In 1999 the absolute frequency measurement resulted in a relative
uncertainty of 1.8 parts in
$10^{14}$~\cite{KolachevskyNiering2000}. In 2003 this measurement
has been repeated and the results of both campaigns are shown in
fig.~\ref{kolachevskyfig9}. In both cases a transportable Cs
atomic fountain clock (FOM) from LNM-SYRTE
Paris~\cite{KolachevskySantarelli1999} has been transported to our
lab at Garching. Its accuracy has been evaluated to $8\times
10^{-16}$, but during the experiments a verification at the level
of $2\times10^{-15}$ only has been performed which is still one
order of magnitude better than required for the $1S$\,--\,$2S$
transition. Unfortunately, the uncertainty of the 2003 measurement
remained nearly the same due to an excessive day-to-day scatter.

The measurements allowed not only to determine the absolute
frequency of the transition, but also to set an upper limit of the
difference of $(-29\pm 57)$~Hz between the measurements that are
44 months apart. This is equivalent to a fractional time variation
of the ratio $f_\textrm{H}/f_\textrm{Cs}$ equal to $(-3.2\pm6.3)
\times 10^{-15}$ yr$^{-1}$, where the ground state hyperfine
splitting of Cs-133, which is used as a reference in these
measurements, is given by $f_\textrm{Cs}$. This limit on the
temporal variation of the absolute optical frequency opens the
possibility to derive an upper limit of the variation of $\alpha$
as detailed in the next section.

\section{High-precision laboratory measurements and variation of the fine structure constant} \label{kolachevskysec2}

So far all laboratory measurements of the drift rates of
fundamental constants are based on comparisons of electromagnetic
transitions that depend in a different way on these constants. The
non-relativistic scalings of gross-, fine- and hyperfine
transitions in atoms, ions and molecules are summarized in
table~\ref{kolachtab1}. A first order theory is sufficient here,
as none of the drifts have been detected yet with small relative
uncertainty. To evaluate the possible drift of $\alpha$ one
measures a frequency ratio of two transitions. Pioneering
astrophysical
measurements~\cite{KolachevskySav56,KolachevskyMinkowski56} used
that method which is now called the ``alkali-doublet method''.

\begin{table}[h]
\begin{center}
\begin{tabular}{|l@{\hspace{2ex}\ }|l@{\hspace{2ex}\ }|r|}
 \hline \hline \rule{-5pt}{3ex}
Sample&Transition&     Scaling factor \\
   \hline
Atom, ion&&\\
&gross structure& $Ry$\\
&fine structure & $\alpha^2 Ry$\\
&hyperfine structure & $g_\textrm{nucl}(\mu_N/\mu_B)\alpha^2Ry $\\
\hline
Molecule &&\\
&gross structure& $Ry$\\
&vibration structure & $(m_e/m_p)^{1/2} Ry$\\
&rotatoinal structure & $(m_e/m_p) Ry$\\
\hline \hline
 \end{tabular}
  \caption{{Scaling factors for different atomic systems in non-relativistic approximation.
  Here $Ry$ is the Rydberg constant in hertz, $g_\textrm{nucl}$ is the nuclear $g$-factor, $\mu_N$ and
  $\mu_B$ -- nuclear and Bohr magnetons respectively, $m_e$ and $m_p$ -- electron and proton mass respectively.
  In the relativistic case, i.e. for heavier atoms, it is necessary to multiply the scalings with relativistic
  correction $F_\textrm{rel}(Z\alpha)$ that depends only on the fine structure constant and may be determined
  from relativistic Hartree-Fock calculations.}}
  \label{kolachtab1}
  \end{center}
\end{table}

In a real situation the values in that table need to be multiplied
with a relativistic correction $F_\textrm{rel}(Z\alpha)$ that
depends only on the fine structure constant and may be determined
from relativistic Hartree-Fock calculations. For hyperfine
transitions in alkali atoms there exists an approximate expression
for the relativistic correction called the Casimir
correction~\cite{KolachevskyCasimir1963} which reads as
\begin{equation}
  F_{\textrm{rel}}(Z\alpha) = \frac{3}{\lambda(4\lambda^2-1)}\,, \qquad
  \textrm{where} \qquad  \lambda \equiv \sqrt{1-(Z\alpha)^2}\,.
\label{kolach_eq11}
\end{equation}
For heavy atomic systems the correction to the hyperfine levels
$F_\textrm{rel}(Z\alpha)$ differs significantly from 1 (e.g.
$F_\textrm{rel}$=1.39 for Cs) so that the sensitivity for $\alpha$
variations may be expressed as:
\begin{equation}
  L^{\textrm{(HFS)}}_\alpha \equiv \alpha\frac{\partial}{\partial\alpha}\ln[F_\textrm{rel}(Z\alpha)]=(Z\alpha)^2\frac{12\lambda^2-1}{\lambda^2(4\lambda^2-1)}
  \label{kolach_eq12}\,.
\end{equation}
For the Cs ground state hyperfine splitting this equals
$L^{\textrm{(HFS)}}_\alpha(\textrm{Cs})\approx0.8$. As only ratios
of frequencies can be determined in a real measurement, the
sensitivity of any experiment will be given by the ratio of the
sensitivity functions of the involved frequencies. Therefore the
sensitivity function defined above is meaningless until it is
referenced to another one.

For optical transition frequencies $f^{\rm{(opt)}}$ no
approximation such as the Casimir correction exists that would be
useful for deriving the leading order dependence on the fine
structure constant. For this reason relativistic Hartree-Fock
calculations have been used. V.A.~Dzuba and co-workers have
expressed the results of their calculation in terms of two
parameters $q_1$ and $q_2$ according to:
\begin{equation}
  f^{\rm{(opt)}}= f^{\rm{(opt)}}_0
  +q_1\left[\left(\frac{\alpha}{\alpha_0}\right)^2-1\right]
  +q_2\left[\left(\frac{\alpha}{\alpha_0}\right)^4-1\right].
  \label{kolach_eq13}
\end{equation}
Here $f^{\rm{(opt)}}_0$ and $\alpha_0$ are the present day (or
laboratory) values of the optical transition frequency and the
fine structure constant respectively. This equation was used to
describe quasar absorption spectra but may be used for laboratory
measurements in which case $f^{\rm{(opt)}}_0$ and $\alpha_0$ are
also laboratory values but at different times. Results for the
parameters $q_1$ and $q_2$ for various atoms and ions including
some important optical clock transitions are published
in~\cite{KolachevskyDzuba1999,KolachevskyDzuba2000}. Only even
powers in $\alpha$ enter the expansion (\ref{kolach_eq13}) because
the relativistic correction is proportional to $\sqrt{m_e^2+p^2}$,
which contains even powers of electron momentum $p\sim Z\alpha$.
Re-expressing Dzuba's notation in terms of the relativistic
correction introduced above yields:
\begin{equation} \label{kolach_eq14}
  L^{\textrm{(opt)}}_\alpha \equiv \alpha \frac{\partial}{\partial \alpha} \ln F_{\rm{rel}}(Z\alpha) =
  \frac{2q_1+4q_2}{f^{\rm{(opt)}}_0}\,.
\end{equation}
Table~\ref{kolachtab2} lists a few values of this quantity adapted
from refs.~\cite{KolachevskyDzuba1999,KolachevskyDzuba2000} that
are relevant for metrological transitions. Note, that for these
calculations the value of $Ry$ has been assumed to be fixed which
imposes a constrain on the value of the product
$m_ec^2\alpha^2/h$. Another way of interpreting this is by picking
$Ry$ as the unit of frequency. Using the same unit for all
frequencies it will eventually drop out of all measurable
quantities as only frequency ratios can be determined in practice.
This will become more clear from the further analysis.
\begin{table}[t!]
\begin{center}
\begin{tabular}{|r|c|c|c@{\hspace{1ex}\ }|r|}
\hline \hline \rule{-5pt}{3ex}
  $Z$   &    {Atom}    &   \qquad Transition \ \ \ \ \ \    &$\lambda$ [nm]& $L^{(opt)}_{\alpha}$   \\
\hline \rule{-5pt}{3ex}
 1& H      &    $\,1\textrm{s}\ \textrm{S}_{1/2}(F=1,m_F=\pm1 )\rightarrow 2\textrm{s} \  \textrm{S}_{1/2}(F'=1,m'_F=\pm1 )$   &121 & 0\\
 20&Ca&${^1\textrm{S}}_{0}(m_J=0)\rightarrow{^3\textrm{P}}_{1}(m_J=0)$&657&0.03\\
 49&In$^+$&$\ \,5\textrm{s}^2\ ^1\textrm{S}_{0}\rightarrow5\textrm{s}5\textrm{p}\ {^3\textrm{P}}_{0} $&237&0.21\\
 70&Yb$^+$ &     $\,6\textrm{s}\ ^2\textrm{S}_{1/2}(F=0)\rightarrow5\textrm{d}\ {^2\textrm{D}}_{3/2}(F=2)$ &435&0.9\\
 80&Hg$^+$ &  $\,5\textrm{d}^{10}6\textrm{s} \ ^2\textrm{S}_{1/2} (F=0) \rightarrow  5\textrm{d}^9 6\textrm{s}^2\ ^2\textrm{D}_{5/2}(F'=2, m'_F=0)$   &282& $-3.2$\\
  \hline
  \hline
  \end{tabular}
  \end{center}
  \caption{ { Sensitivity of relativistic corrections $F_{\rm{rel}}(Z\alpha)$
  to $\alpha$ for some atomic transitions according to~\cite{KolachevskyDzuba1999,KolachevskyDzuba2000}.}}
   \label{kolachtab2}
\end{table}

Comparing optical transitions with different relativistic
corrections became a powerful instrument to set upper limits to
the drift of fundamental constants. This method is widely used in
astrophysical observations (``many-multiplet''
method~\cite{KolachevskyWebb2001}) and in laboratory comparisons.
An elegant realization of this method has been used by A.~Cing\"oz
{\it et al.}~\cite{KolachevskyCingoz2007}, by utilizing different
relativistic corrections of two nearly degenerate levels of
opposite parity in neutral dysprosium. Monitoring the radio
frequency transitions at 3.1\,MHz for $^{163}$Dy and 325\,MHz for
$^{162}$Dy during 8 months only, the authors set a stringent limit
on the drift of the fine-structure constant of $\partial
\ln(\alpha)/\partial t = (-2.7\pm2.6)\times 10^{-15}$\,yr$^{-1}$
without any assumptions about the drift of other constants. In the
next section we will describe how one can deduce a
model-independent restriction to $\dot\alpha$ from different
absolute optical frequency measurements.

\subsection{Upper limit for the drift of the fine structure
constant from optical frequency measurements}\label{kolachsubsec3}

\subsubsection{Absolute frequency measurements}

Thanks to the optical frequency comb the determination of absolute
optical transition frequencies became a simple task, where the
attribute ``absolute'' means that the frequency is measured in
hertz, i.e. in terms of the Cs ground state hyperfine splitting.
For this a ratio like
$f^{\rm{(opt)}}/f^{(\rm{HFS})}_{\textrm{Cs}}$ is determined.
According to table~\ref{kolachtab1} such a ratio depends on two
fundamental constants, $\alpha$ and the Cs nuclear magnetic moment
measured in Bohr magneton's $\mu_\textrm{Cs}/\mu_B$. One may argue
that the latter is not a fundamental quantity, but one has to keep
in mind that the nuclear moment is mostly determined by the strong
interaction. In that sense it measures the strong interaction in
some units. The only difference to the electromagnetic interaction
measured by $\alpha$ is that, lacking a precise model for the Cs
nucleus we are not sure what those units are. For this reason
there are two parameters that need to be determined from an
absolute optical frequency measurement and it is impossible to
disentangle contribution from the drift rate of just one absolute
optical frequency. At the other hand, the task is solvable if one
has more than one absolute frequency measurement at hand under the
condition, that the values $L_\alpha^\textrm{opt}$ are different
for these transitions.

For the general case let's assume that there are $N$ repeated
absolute frequency measurements of corresponding transitions
$\textrm{T}_i$. For each of the transitions one can derive the
relative drift of its absolute frequency
 $b_i$ ($i=1\ldots N$) as well as the corresponding uncertainty
$\sigma_i$ (one standard deviation)
\begin{equation}
  \label{kolach_eq15}
 \frac{\partial}{\partial t}
  \ln\frac{f^{(\rm{HFS})}_{\textrm{Cs}}}{f^{\rm{(opt)}}_{\textrm{T}_i}}=
  b_i\pm\sigma_i\,.
\end{equation}
One can rewrite (\ref{kolach_eq15}) using the results from table
\ref{kolachtab2} and eqn. (\ref{kolach_eq14})
\begin{eqnarray}
  \label{kolach_eq16}
  \frac{\partial}{\partial t}
  \ln\frac{f^{({\mathrm{HFS}})}_{\mathrm{Cs}}}{f^{{\mathrm{(opt)}}}_{\mathrm{T}_i}}
  &=&
  \frac{\partial}{\partial t}\left[\ln\left(\frac{\mu_\textrm{Cs}}{\mu_B}\right)+(2+L^{\rm{(HFS)}}_\alpha({\textrm{Cs}})- L^{\rm{(opt)}}_\alpha({\textrm{T}_i}))\ln\alpha\right] = \nonumber\\
  &=&y+A_i\,x\,,
\end{eqnarray}
where we introduced the definitions
$y\equiv\partial\ln({\mu_\textrm{Cs}}/{\mu_B})/\partial t$ and
$x\equiv\partial\ln(\alpha)/\partial t$ respectively. The
coefficient $A_i$ incorporate sensitivities of the corresponding
relativistic corrections $L_\alpha$ as well as the $\alpha^2$
scaling for hyperfine transitions. Thus, experiments relate $x$
and $y$ to measured values $b_i$ with uncertainties $\sigma_i$
 through:
\begin{equation}
  \label{kolach_eq17}
  y=A_i\,x+b_i\pm\sigma_i\,.
\end{equation}
Let's assume further that the measured data follows a Gaussian
distribution $P(x,y)$:
\begin{equation}\label{kolach_eq18}
P(x,y)\propto e^{-\frac{1}{2}R^2(x,y)}\,, \  \ \textrm{where} \ \
R^2=\sum_i\frac{1}{\sigma_i^2}(y-A_i\, x-b_i)^2\,.
\end{equation}
The expectation values for the relative drift rates $x$ and $y$
are determined by the maximum likelihood method corresponding to
the minimum of $R^2(x,y)$:
\begin{eqnarray}\label{kolach_eq19}
\frac{\partial R^2}{\partial
x}&=&-2\sum\frac{1}{\sigma_i^2}(y-A_i\,
x-b_i)A_i=0\\
\frac{\partial R^2}{\partial
y}&=&-2\sum\frac{1}{\sigma_i^2}(y-A_i\, x-b_i)=0\, .\nonumber
\end{eqnarray}
With the definitions $B_1\equiv\sum{1}/{\sigma^2_i}$,
$B_2\equiv\sum{A^2_i}/{\sigma^2_i}$,
$B_3\equiv\sum{b^2_i}/{\sigma^2_i}$,
$B_4\equiv\sum{A_i}/{\sigma^2_i}$,
$B_5\equiv\sum{b_i}/{\sigma^2_i}$,
$B_6\equiv\sum{A_i\,b_i}/{\sigma^2_i}$ we can solve system
(\ref{kolach_eq19}) for  $x$ and  $y$ and obtain expressions for
the expectation values:
\begin{equation} \label{kolach_eq20}
  \langle x\rangle=\frac{B_4\,B_5-B_1\,B_6}{B_1\,B_2-B^2_4}\,,\qquad
  \langle y\rangle=\frac{B_2\,B_5-B_4\,B_6}{B_1\,B_2-B^2_4}\,.
\end{equation}
The standard deviation for $\langle x \rangle$ can be calculated
from the integral:
\begin{equation}\label{kolach_eq21}
  \int_{-\infty}^{+\infty}e^{-\frac{1}{2}R^2(x,y)}dy\propto
  \exp\left[\frac{(B_5+x\,B_4)^2-B_1(B_3+x\,(x\,B_2+2B_6))}{2B_1}\right]\,.
\end{equation}
Rewriting the exponent
\begin{equation} \label{kolach_eq22}
  \exp\left[-\frac{(x-\langle x \rangle)^2}{2\sigma^2_x}+const_x\right]\,,
\end{equation}
one gets the standard deviation for $x$
\begin{equation} \label{kolach_eq23}
  \sigma_x=\sqrt{\frac{B_1}{B_1\,B_2-B^2_4}}
\end{equation}
and, similarly, for $y$
\begin{equation} \label{kolach_eq24}
  \sigma_y=\sqrt{\frac{B_2}{B_1\,B_2-B^2_4}}\,.
\end{equation}
The evaluation may be represented graphically as on the left hand
side of fig.~\ref{kolachevskyfig2}.

As an example consider the results of the absolute frequency
measurements of the $1S$\,--\,$2S$ transition in atomic hydrogen
(data taken during 2001-2003 at our
lab~\cite{KolachevskyFischer2004}):
\begin{eqnarray}\label{kolach_eq25}
  - \frac{\partial}{\partial t} \ln\frac{f^{\rm{(opt)}}_{\textrm{H}}}{f^{(\rm{HFS})}_{\textrm{Cs}}}
  =\frac{\partial}{\partial t}\left[\ln\left(\frac{\mu_\textrm{Cs}}{\mu_B}\right)+(2+0.8)\ln\alpha\right] = \nonumber\\
  =y+2.8x=(3.2\pm6.4) \times 10^{-15}\ \ \textrm{yr}^{-1}\,,
\end{eqnarray}
the frequency measurement of the electric quadrupole transition
$5\textrm{d}^{10}6\textrm{s} \ ^2\textrm{S}_{1/2}\
(F=0)$\,--\,$5\textrm{d}^9 6\textrm{s}^2\ ^2\textrm{D}_{5/2} \
(F'=2, m'_F=0)$ at  $\lambda=282$\,nm  in a single laser cooled
$^{199}$Hg${^+}$ ion (data taken during 2000-2006 at NIST
\cite{KolachevskyFortier2007}):
\begin{eqnarray}\label{kolach_eq26}
  - \frac{\partial}{\partial t}\ln\frac{f^{\rm{(opt)}}_{\textrm{Hg}}}{f^{(\rm{HFS})}_{\textrm{Cs}}}
  =\frac{\partial}{\partial t}\left[\ln\left(\frac{\mu_\textrm{Cs}}{\mu_B}\right)+(2+0.8+3.2)\ln\alpha\right] = \nonumber\\
  =y+6x=(-0.37\pm0.39)\times 10^{-15}\ \ \textrm{yr}^{-1}\,,
\end{eqnarray}
and the frequency measurement of the $6\textrm{s}\
{^2}\textrm{S}_{1/2}(F=0)$\,--\,\,$6\textrm{s}\
{^2}\textrm{D}_{3/2}(F=3)$ electric quadrupole transition at
$\lambda = 436$\,nm of a single trapped and laser cooled
$^{171}$Yb$^+$ ion (data taken during 2000-2006 at PTB
\cite{KolachevskyPeik2004,KolachevskyPeik2006}):
\begin{eqnarray}\label{kolach_eq27}
  - \frac{\partial}{\partial t} \ln\frac{f^{\rm{(opt)}}_{\textrm{Yb}}}{f^{(\rm{HFS})}_{\textrm{Cs}}}
  =\frac{\partial}{\partial t}\left[\ln\left(\frac{\mu_\textrm{Cs}}{\mu_B}\right)+(2+0.8-0.9)\ln\alpha\right] = \nonumber \\
  =y+1.9x=(0.78\pm1.4)\times 10^{-15}\ \ \textrm{yr}^{-1}.
\end{eqnarray}

Using the experimental data from eqns. (\ref{kolach_eq25}),
(\ref{kolach_eq26}),  (\ref{kolach_eq27}) and expressions
(\ref{kolach_eq20}), (\ref{kolach_eq23}), (\ref{kolach_eq24})
stringent restrictions for fractional variations of the
fundamental constants can be derived
\cite{KolachevskyFortier2007}:
\begin{eqnarray}  \label{kolach_eq28}
  x&=&\frac{\partial}{\partial{t}} \ln\alpha=(-0.31\pm0.35)\times10^{-15}\,\textrm{yr}^{-1}\,,\\
  y&=&\frac{\partial}{\partial{t}}\ln\frac{\mu_\textrm{Cs}}{\mu_B}=(1.5\pm2.0)\times10^{-15}\,\textrm{yr}^{-1}\,.  \label{kolach_eq29}
\end{eqnarray}
This result does not use any assumption on correlation of
fundamental constants. It is an important advantage of the method
opening the possibility to test some extensions of the grand
unification theories  where the strong, weak and electromagnetic
coupling constants are expected to merge for higher energies. The
drifts  (if existing) of corresponding coupling constants should
be correlated; one can even derive a relation of the drift rates
of hadron masses and nuclear $g$-factors that are determined by
the strong interaction, and the relative drift rate of the fine
structure constant: $\Delta m_p/m_p \approx \Delta
g_\textrm{nucl}/g_\textrm{nucl} \approx \pm 35 \Delta
\alpha/\alpha$~\cite{KolachevskyCalmet2002a,KolachevskyCalmet2002b}.
Of course, the theory can be tested only if nonzero drift rate is
detected as the relation holds even if none of the constants is
actually drifting.

\subsubsection{Coupling to gravity}

Besides setting an upper limit to a variation of $\alpha$,
repeated absolute frequency measurements deliver important
information about the coupling between gravity and other
fundamental interactions. Since 2005 the $^1S_0$\,-\,$^3P_0$ clock
transition frequency in $^{87}$Sr has been measured relative to
the Cs standard at three laboratories in Paris, Boulder and Tokyo
with gradually improving accuracy~\cite{KolachevskyBlatt2008}. In
these experiments Sr atoms are placed at the minima of a periodic
optical potential (an ``optical lattice'') tuned to a selected
wavelength~\cite{KolachevskyKatori2003} which prevents the clock
transition to be shifted by the optical potential. This generates
extremely narrow (down to 2\,Hz at 698\,nm) optical resonances in
a large ensemble of atoms. Results agree at a level of $10^{-15}$
so that this type of optical clocks among the most accurate.

Besides improvement of the null-result
$\dot\alpha/\alpha=(-3.3\pm3.0)\times10^{-16}\,\textrm{yr}^{-1}$,
an upper limit for coupling between gravity and other fundamental
interactions was set. The Earth moves on an elliptic orbit in a
varying solar gravitational potential with a fractional variation
of up to $3.3\times 10^{-10}$. If the coupling between $\alpha$
and the variation of the gravitational potential $\Delta U(t)$ is
assumed to be of the form
\begin{equation} \label{kolach_eq30}
\frac{\delta\alpha}{\alpha}=k_\alpha\frac{\Delta U(t)}{c^2}\,,
\end{equation}
where $k_\alpha$ is the coupling constant, the authors
of~\cite{KolachevskyBlatt2008} have set a restriction of
\begin{equation} \label{kolach_eq31}
k_\alpha=(2.5\pm3.1)\times10^{-6}.
\end{equation}
The same coupling constant has previously been limited with
transitions in dysprosium (see also the beginning of
sec.~\ref{kolachevskysec2}) with approximately half the
sensitivity~\cite{KolachevskyFerrell2007}.

Of course, the sensitivity of this type of measurement depends on
fractional variation of the gravitational potential which is
rather small for the Sun-Earth system. The idea of performing
atomic clock frequency comparisons at larger values of $\Delta
U/c^2$ was considered previously within the {\it{SpaceTime}}
satellite mission~\cite{KolachevskyLammerzahl2002} in which a
fly-by maneuver near Jupiter could increase $\Delta U/c^2$ to
$5\times10^{-7}$.

\subsubsection{Direct comparison of optical frequencies}

Improving the accuracy of absolute optical frequency measurements
one eventually encounters the limit set by the stability and
accuracy of the best Cs clocks (see fig.~\ref{kolachevskyfig1}).
The accuracy of the currently best state-of-art fountain clocks is
around $5\times10^{-16}$~\cite{KolachevskyBize2005}. As the
frequency combs are not limiting at this level (see e.g.
\cite{KolachevskyZimmermann2004}) direct comparison of two optical
clocks (cec.~\ref{kolachevskysubsec223}) can provide improved data
if both of these clocks are more accurate than the Cs clocks would
be.

Indeed at  NIST (Boulder, USA) two of these clocks are available
that are based on optical clock transition frequencies in Hg$^+$
and Al$^+$~\cite{KolachevskyRosenband2008}. The frequency ratio
measured with the help of a frequency comb has a relative
uncertainty of only $5.2\times10^{-17}$ which is an order of
magnitude smaller than for any absolute frequency measurement.
This breakthrough became possible after implementation of new
concepts in probing of clock transitions in cold ions, development
of narrow-band lasers and progress in optical frequency transfer.
Comparison of a highly relativistic system of Hg$^+$ (the
sensitivity to $\alpha$ valuation is
$L^\mathrm{opt}_\alpha(\textrm{Hg}^+)=-3.2$) with a nearly
non-relativistic system of Al$^+$
($L^\mathrm{opt}_\alpha(\textrm{Al}^+)=+0.008$)) over a time
interval of only one year allowed to derive a restriction to the
variation of $\alpha$ of
\begin{equation}\label{kolach_eq32}
\dot\alpha/\alpha=(-1.6\pm2.3)\times10^{-17}\,\textrm{yr}^{-1}\,.
\end{equation}
This is the lowest limit obtained yet of any type of measurement
and is consistent with zero. On top of that the result is
model-independent and opens the possibility to draw conclusions
about variations of other fundamental constants when combined with
other types of experiments.

\section{Frequency combs for astrophysics}\label{kolachevskysec3}

The laser frequency comb turned out to be not only an
indispensable element in laboratory optical frequency
measurements, but is also in the process of proofing itself useful
for astronomical observations. It has been demonstrated recently
that a frequency comb with a resolvable large mode spacing can be
used as an accurate calibration tool for high-resolution
spectrographs. This is particular interesting for astrophysical
applications~\cite{KolachevskySteinmetz2008}. An accurate
frequency axis for spectrometers is required for a number of
sensitive fundamental measurements like testing the drift of
redshifts of different astrophysical
objects~\cite{KolachevskySandage1962}, the search for extrasolar
planets by the reflex Doppler
motion~\cite{KolachevskyMayor1995,KolachevskyMarcy1996,KolachevskyLovis2006}
as well as the search for cosmological variations of fundamental
constants~\cite{KolachevskyBahcall1965,KolachevskyThompson1975,KolachevskyWebb1999}.

If Doppler shifts on the order of 1\,cm\,s$^{-1}$ ($3 \times
10^{-11}c$) could be measured the presumed acceleration of the
cosmic expansion could be verified in real time in a largely model
independent way, i.e. without assuming the validity of general
relativity~\cite{KolachevskySteinmetz2008}. For a typical
high-resolution spectrometer used in astrophysics such a
resolution corresponds to a physical size of one silicon atom on a
CCD substrate. This indicates that only with the statistics of a
very large number of calibration lines the sensitivity can be
achieved at the condition that the systematics are under control
at the same level.

\begin{figure}[t!]
\begin{center}
\includegraphics[width=1.0\textwidth]{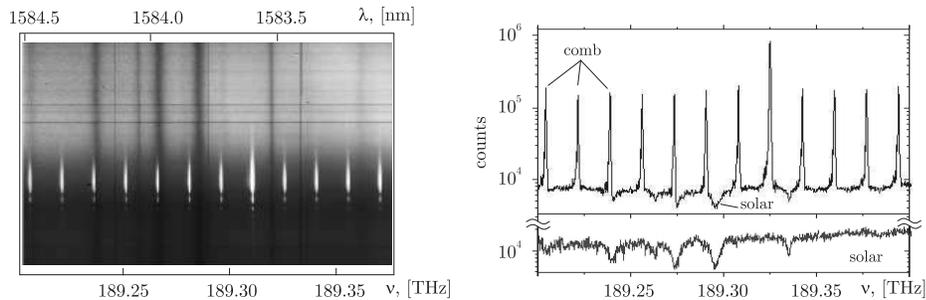}
\caption[]{{(left) --- a CCD image of a fragment of the solar
photosphere spectrum (dark Fraunhofer lines) overlaid with a
frequency comb with 15\,GHz mode spacing (bright regular lines).
(right) --- horizontal cuts through the CCD image that contain the
frequency comb and solar spectrum (top) and the solar spectrum
only (bottom).}} \label{kolachevskyfig10}
\end{center}
\end{figure}
In 2008 a first implementation of the laser frequency comb as a
calibration tool for the German Vacuum Tower Telescope
(VTT)~\cite{KolachevskySchroter1985} has been
demonstrated~\cite{KolachevskySteinmetz2008}. The very high
resolution of the VTT (0.8\,GHz) is still too low to resolve
individual modes of the erbium fiber laser frequency comb with
$\omega_r=2\pi \times 250$~MHz used for its calibration. Filtering
of the desirable frequency comb modes by a external Fabry-P\'erot
cavity was suggested as one possible solution (e.g.
\cite{KolachevskyLi2008}). The Fabry-P\'erot cavity used
in~\cite{KolachevskySteinmetz2008} has a free spectral range of $m
\omega_r$ where the integer $m$ can be set between 4 and 60. This
cavity interferometrically suppresses all modes generated by the
laser except every $m$th. The resulting well resolved comb is used
to illuminate the spectrograph slit.

A CCD image of the fragment of the solar spectrum with atmospheric
absorption lines is presented in fig.~\ref{kolachevskyfig10}. The
filtered frequency comb radiation was overlaid with the output of
the telescope and sent to the VTT spectrometer. The comb modes
were stabilized to a Rb atomic clock. One can recognize the
resolved filtered comb modes used as frequency markers separated
by 15\,GHz intervals ($m=60$) super imposed on the solar spectrum.
Even for this very first demonstration a calibration uncertainty
of only 9\,m\,s$^{-1}$ (root mean square) in cosmic velocity units
was achieved which compares every well with the uncertainties of
traditional calibration techniques. By increasing the number of
modes (up to $10^4$) by a spectrally broader frequency comb, it is
feasible to reduce the statistical uncertainty to the desirable
1\,cm\,s$^{-1}$ level. The approach also opens possibilities to
analyze the systematic uncertainties of the spectrograph and
remove their contribution. Implementation of frequency combs for
astrophysics allows to observe small variations of spectral lines
on a large time scale referenced directly to the SI unit of hertz.

\section*{Conclusions}

Table \ref{kolachtab3} summarizes results of recent laboratory
measurements aiming at the search for a time varying $\alpha$(see
also fig.~\ref{kolachevskyfig2}, right). Combinations with other
precision laboratory measurements like comparisons of fountain
clocks (see e.g. \cite{KolachevskyMarion2003}) or precision
molecular spectroscopy \cite{KolachevskyKlein2005} deliver
important information on the variation of reduced magnetic moments
and the electron-to-proton mass ratio. The field is rapidly
evolving and we included only a few results most relevant to the
topic reviewed.

\begin{table}[h!]
\begin{center}
\begin{tabular}{|l@{\hspace{1ex}\ }|c@{\hspace{3ex}\ }|@{\hspace{3ex}\ }c@{\hspace{3ex}\ }|c|c|}
  \hline \hline \rule{-5pt}{3ex}
  Year & Atomic samples & $\dot\alpha/\alpha$, yr$^{-1}$ & Method & Ref. \\
  \hline \rule{-5pt}{3ex}
 2004 & H, Yb$^{+}$, Hg$^{+}$ & $(-0.9\pm2.9)\times10^{-15}$ & absolute frequency & \cite{KolachevskyFischer2004} \\ \rule{-5pt}{1ex}
 2004 &H, Yb$^{+}$, Hg$^{+}$ & $(-0.3\pm2.0)\times10^{-15}$ & absolute frequency &
 \cite{KolachevskyPeik2004}\\ \rule{-5pt}{1ex}
 2006 &  Yb$^{+}$, Hg$^{+}$ & $(-2.6\pm3.9)\times10^{-16}$ & absolute frequency& \cite{KolachevskyPeik2006}
 \\ \rule{-5pt}{1ex}
 2007 &  Dy & $(-2.7\pm2.6)\times10^{-15}$ & rf transition & \cite{KolachevskyCingoz2007}
 \\ \rule{-5pt}{1ex}
 2008 &  Sr, H, Yb$^{+}$, Hg$^{+}$ & $(-3.3\pm3.0)\times10^{-16}$ & absolute frequency & \cite{KolachevskyBlatt2008}
 \\ \rule{-5pt}{1ex}
 2008 &  Hg$^{+}$, Al$^{+}$ & $(-1.6\pm2.3)\times10^{-17}$ & direct comparison& \cite{KolachevskyRosenband2008} \\
  \hline\hline
\end{tabular}
\caption{Model-independent restrictions of the variation of the
fine structure constant $\alpha$ from laboratory measurements.
Results
of~\cite{KolachevskyFischer2004,KolachevskyPeik2004,KolachevskyPeik2006,KolachevskyBlatt2008}
were obtained by absolute optical frequency measurements, the
result~\cite{KolachevskyRosenband2008} was obtained by direct
comparison of two optical frequencies with the help of a frequency
comb, while in~\cite{KolachevskyCingoz2007} a radio-frequency
transition between highly-excited nearly degenerative levels has
been measured. }
  \label{kolachtab3}
  \end{center}
\end{table}

As seen from the table, repeated frequency measurements in cold
atoms, ions and molecules allow to set stringent restrictions on
the variation of fundamental constants. At a moment the
sensitivity of these methods resides at a level of
$\dot\alpha/\alpha\sim10^{-17}$\,yr$^{-1}$ which is the lowest
model-independent restriction at the present epoch. A further
increase of the sensitivity is expected due to improvements of
frequency measurements, the increasing observation time interval,
and the increase of the number of atomic samples under study. On
the other hand, frequency combs open perspectives for improving
the accuracy of astrophysical observations and potentially push
forward the sensitivity of astrophysical tests for a variation of
$\alpha$ billions of years ago.

\section*{Acknowledgements} J.A. is supported by EU
Marie Curie fellowship. N.K. acknowledges support from MPG and RF
Presidential grant MD-887.2008.2.

\end{document}